\documentclass[11pt]{article}
\usepackage{graphicx}

\newcommand{\BABARPubYear}    {07}

\newcommand{\BABARConfNumber} {007}
\newcommand{\SLACPubNumber} {12731}

\def\GeV{\;\mbox{GeV}}

\def\GeVcc{\;\mbox{GeV}/c^2}

\def\MeVcc{\;\mbox{MeV}/c^2}

\def\dE   {\Delta E}
\def\dz   {\Delta z}
\def\mes  {M_{\mbox{\scriptsize ES}} }

\def\Bp {B^+}
\def\Bz {B^0}

\def\bsg    {b\to s\gamma}

\def\Kz    {\ensuremath{K^{0}}\xspace}

\def\de        {\ensuremath {\Delta E}}

\def\itypeone 	{\ensuremath{\Bz \to \pipi \gamma}}
\def\itypetwo	{\ensuremath{\Bp \to \pip \piz \gamma}}
\def\itypethree	{\ensuremath{\Bp \to \pipi \pip \gamma}}
\def\itypefour	{\ensuremath{\Bz \to \pipi \piz \gamma}}
\def\itypefive	{\ensuremath{\Bz \to \pipi \pipi \gamma}}
\def\itypesix	{\ensuremath{\Bp \to \pipi \pip \piz \gamma}}
\def\itypeseven	{\ensuremath{\Bp \to \pip \eta \gamma}}

\input babarsym
 
\long\def\inst#1{\par\nobreak\kern 4pt\nobreak
    {\it #1}\par\vskip 10pt plus 3pt minus 3pt}

\begin{document}
{\pagestyle{empty}

\begin{flushright}
\babar-CONF-\BABARPubYear/\BABARConfNumber \\
SLAC-PUB-\SLACPubNumber \\
July 2007 \\
\end{flushright}

\par\vskip 5cm

\begin{center}
\Large \bf \boldmath
Evidence for \btodgam\ Transitions From a Sum of
Exclusive Final States in the Hadronic Final State
Mass Range $1.0 \gevcc< M(X_d)<1.8 \gevcc$.
\end{center}
\bigskip

\begin{center}
\large The \babar\ Collaboration\\
\mbox{ }\\
\today
\end{center}
\bigskip \bigskip

\begin{center}
\large \bf Abstract
\end{center}
We present preliminary results of a search for $B\to X_d\gamma$ decays  
with a hadronic mass $1.0\GeVcc<M(X_d)<1.8\GeVcc$. We consider seven final states 
with up to four charged pions and one neutral pion or $\eta$, which correspond to about 
$50\%$ of the total $X_d$ fragmentation in this mass range. Based on a sample of 
383 million $B\bar{B}$ events collected by the BaBar experiment at PEP-II, we measure 
a partial branching fraction 
$\sum\nolimits^{7}_{X_d=1}\BR(B\to X_d\gamma)|_{(1.0\gevcc<M(X_d)<1.8\,\mathrm{GeV}/c^2)}= (3.1\pm0.9 ^{+0.6}_{-0.5} \pm0.5)\cdot 10^{-6},$ 
where the uncertainties are statistical, systematic and model-dependent respectively. 
\vspace*{8ex}

\begin{center}
Contributed to the 
XXIII$^{\rm rd}$ International Symposium on Lepton and Photon Interactions at High~Energies, 8/13 -- 8/18/2007, Daegu, Korea
\end{center}

\vspace{1.0cm}
\begin{center}
{\em Stanford Linear Accelerator Center, Stanford University, 
Stanford, CA 94309} \\ \vspace{0.1cm}\hrule\vspace{0.1cm}
Work supported in part by Department of Energy contract DE-AC03-76SF00515.
\end{center}

\newpage
} 

\begin{center}
\small

The \babar\ Collaboration,
\bigskip

%
{B.~Aubert,}
{M.~Bona,}
{D.~Boutigny,}
{Y.~Karyotakis,}
{J.~P.~Lees,}
{V.~Poireau,}
{X.~Prudent,}
{V.~Tisserand,}
{A.~Zghiche}
\inst{Laboratoire de Physique des Particules, IN2P3/CNRS et Universit\'e de Savoie, F-74941 Annecy-Le-Vieux, France }
{J.~Garra~Tico,}
{E.~Grauges}
\inst{Universitat de Barcelona, Facultat de Fisica, Departament ECM, E-08028 Barcelona, Spain }
% ...

\end{center}\newpage


\section{INTRODUCTION}
\label{sec:Introduction}
\begin{figure}[!h]
\begin{center}
\includegraphics[width=250pt]{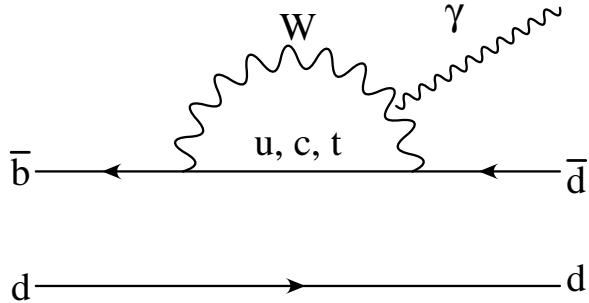}
\caption{Feynman diagram for a $\bar{b}$ \to $\bar{d}\g$ transition}
\label{fig:feyndiag} 
\end{center}
\end{figure}

We present an experimental study of the rare decays $B\to X_d\gamma$. Within the 
standard model of particle physics (SM), these flavor-changing-neutral-current (FCNC) 
$b\to d\gamma$ transitions are forbidden at tree level; the leading-order processes 
are one-loop electroweak penguin diagrams (see Figure~\ref{fig:feyndiag}), 
where the top quark is the dominant virtual quark contribution. 
In the context of theories beyond the SM, new virtual particles may appear 
in the loop, which could lead to measurable effects on experimental 
observables such as branching fractions and $CP$ asymmetries~\cite{bsm}.

Previous measurements of the exclusive decays $B^{+} \to \rho^{+}\gamma$, 
$B^{0} \to \rho^{0}\gamma$, and $B^{0} \to \omega\gamma$ by the 
Belle~\cite{bellerhog} and \babar~\cite{babarrhog} 
experiments found branching fractions of $\mathcal{O}(10^{-6})$,  
in good agreement with SM predictions~\cite{SM}. 
In combination with the well-measured $\B\to K^*\gamma$ 
branching fractions they also yielded measurements of the ratio of Cabbibo-Kobayashi-Maskawa 
(CKM) matrix elements $|V_{td}/V_{ts}|$ which are in good agreement with results 
independently obtained from the ratio of $B_d^0$ and $B_s^0$ mixing frequencies~\cite{bsmixing}.

The experimental and theoretical uncertainties associated with the exclusive 
measurements are still large and there is a strong motivation to extend the experimental 
study to additional final states and different regions of the hadronic mass spectrum, with a 
measurement of the inclusive $b\to d\gamma$ transition rate as the ultimate goal.
In this note we report the first study of $\btodgam$ transitions using 
a sum of seven exclusive $X_d \gamma$ final states in the previously unaccessed hadronic mass range 
$1.0\gevcc<M(X_d)<1.8\,\mathrm{GeV}/c^2$.

\section{THE \babar\ DETECTOR}
\label{sec:Detector}
The data used in this analysis were collected with the \babar\ detector
at the \pep2\ asymmetric--energy $\epem$ storage ring. Charged particle
trajectories are measured using a five-layer silicon vertex
tracker and a 40-layer drift chamber in a 1.5-T
magnetic field. Photons and electrons are detected in a CsI(Tl) crystal
electromagnetic calorimeter (EMC) with photon energy resolution
$\sigma_E / E = 0.023 (E/\mathrm{GeV})^{-1/4} \oplus 0.019$. A ring-imaging
Cherenkov detector (DIRC) is used for charged-particle identification.
In order to identify muons, the magnetic flux return is instrumented with
resistive plate chambers and limited streamer tubes. A detailed description
of the detector can be found elsewhere~\cite{ref:babar}.

\section{EVENT RECONSTRUCTION AND SELECTION}
\label{sec:Selection}

We exclusively reconstruct seven \btodgam\ decay modes with up to four charged pions and one 
neutral pion or eta in the final state: \itypeone, \itypetwo, 
\itypethree, \itypefour, \itypefive, \itypesix, \itypeseven\ \cite{conj}. 

An important source of background is due to continuum events ($\ep\en \to q\bar{q}$, 
with $q=u,d,s,c$) that contain a high-energy photon from $\piz$ or $\eta$ decays or 
from initial-state radiation. There is also background from combinatorial $B$ decays. 
These include $\B\rightarrow X_d\piz$ and $\B\rightarrow X_d\eta$ processes that produce a high-energy 
photon in the $\piz$ or $\eta$ decay, as well as $B\rightarrow X_s\gamma$ decays 
where a $\Kpm$ is misidentified as a $\pipm$. 
Background suppression cuts have been optimized  for maximum statistical sensitivity 
$S/\sqrt(S+B)$, where $S$ and $B$ are the rates for signal and background events,  
using Monte Carlo (MC) simulated event samples, and assuming an inclusive 
$b\to d\gamma$ branching fraction of $1.0\times 10^{-5}$.
Of particular concern are backgrounds from \B\ meson decays that resemble the signal.

The photon from the signal $B$ decay is identified as a localized energy deposit in the 
calorimeter with energy $1.15 <E^*_\gamma <3.5\GeV$ in the  center-of-mass (CM) frame. 
The energy deposit must not be associated with any reconstructed charged track, be 
well isolated from other EMC deposits, and meet a number of additional requirements designed 
to eliminate background from hadronic showers and small-angle photon pairs. 
We veto any photons that can form a $\piz$ ($\eta$) 
candidate by association with another 
detected photon of energy greater than 30 (250)~MeV  by requiring that the two--photon invariant mass be outside the range of 
$105\MeVcc$--$155\MeVcc$ ($500\MeVcc$--$590\MeVcc$).

Charged pion candidates are selected from well-reconstructed tracks with a minimum 
momentum in the laboratory frame of $300\mevc$. In order to reduce backgrounds from 
charged kaons produced in $\bsg$ processes, a $\pipm$ selection algorithm is applied, 
combining DIRC information with the energy loss measured in the tracking system. 
At a typical pion energy of 1\gev\ the pion selection efficiency is over 85\%\ 
and the Kaon mis-identification rate is less than 2\%. 

$\piz$ ($\eta$) candidates are formed from pairs of photons with energies greater 
than $20\mev$, with an invariant mass $117 < m_{\gamma\gamma} < 145\MeVcc$ 
($470 < m_{\gamma\gamma} < 620\MeVcc$). 
We require $\piz$ and $\eta$ candidates to have a momentum greater than $300\mevc$. 

The selected $\pipm$, 
$\piz$, $\eta$, and high-energy photon candidates are combined to form $B$ meson 
candidates consistent with one of the seven signal decays. Here, the charged pions 
are combined to form a common vertex, where we require the vertex probability  
be greater than 2\%. The mass of the hadronic system is required to be in the range 
$1.0\GeVcc <M(X_d)< 1.8\GeVcc$. We define 
$\dE \equiv E^*_{B}-E_{\rm beam}^*$, where $E^*_B$ is the CM energy of the $B$-meson 
candidate and $E_{\rm beam}^*$ is the CM beam energy. We also use the beam-energy-substituted 
mass $\mes \equiv\sqrt{ E^{*2}_{\rm beam}-{\mathrm{\vec{p}}}_{B}^{\;*2}}$, where 
${\mathrm{\vec{p}}}_B^{\;*}$ is the CM momentum of the $B$ candidate. Signal events 
are expected to have a $\dE$ distribution centered at zero with a resolution of about 
$30\mev$,  and an $\mes$ distribution centered at the mass of the $B$ meson, $m_B$, with 
a resolution of $3~\mevcc$. We consider candidates in the ranges $-0.3\gev < \dE <0.2 \gev$ 
and $\mes >  5.22\GeVcc$ to incorporate sidebands that allow the combinatorial background 
yields to be extracted by a fit to the data.

Contributions from continuum background processes are reduced by considering only events 
for which the ratio $R_2$ of second-to-zeroth order Fox-Wolfram moments~\cite{fox} is 
less than 0.9.  To further discriminate between the jet-like continuum background and 
the more spherically-symmetric signal events, we compute the angle $\theta_T$ between the 
photon and the thrust axis of the rest of the event (ROE) in the CM frame and require 
$|cos(\theta_T)|<0.8$. The ROE is defined as all the charged tracks and neutral energy 
deposits in the calorimeter that are not used to reconstruct the $B$ candidate.

The quantity $cos(\theta_T)$ and twelve other variables that distinguish between 
signal and continuum events are combined in a neural network (NN): $R'_2$, 
which is $R_2$ in the frame recoiling against the photon momentum, is used to suppress events 
with initial-state radiation. We also compute the $B$-meson production angle 
$\theta_B^*$ with respect to the beam axis in the CM frame, and the Legendre moments 
$L_{i} \equiv \sum_{j} p^{*}_{j}
\cdot|\cos{\theta^{*}_{j}}|^{i}/\sum_{j} p^{*}_{j}$ and 
$\tilde{L}_{i} \equiv \sum_{j} p^{*}_{j}
\cdot|\sin{\theta^{*}_{j}}|^{i}/\sum_{j} p^{*}_{j}$, where 
$p^{*}_j$ and $\theta^{*}_{j}$ are the momentum and angle with respect to a given axis,
respectively, for each particle $j$ in the ROE. We use $L_{1}$ with respect to the ROE 
thrust axis, as well as $L_{2}$, $L_{3}$, $\tilde{L}_{2}$, and $\tilde{L}_{3}$ with 
respect to the high--energy photon direction as NN input. Differences in lepton and kaon 
production between background and \B decays are exploited by including several flavor-tagging 
variables described in \cite{babartag}. 
We select events for which the NN yields an output value greater than 0.83; this corresponds 
to signal and continuum background efficiencies of about $50\%$ and $0.5\%$ respectively.

The expected average number of candidates per selected signal event 
is 1.75. 
In events with multiple candidates the one with the reconstructed 
$\piz (\eta)$ mass closest to the nominal \cite{PDG} is retained. 
For events containing no $\piz (\eta)$ the candidate with the highest vertex 
probability is retained. 

From simulated signal events with $1.0\gevcc<M_{X_d}<1.8\,\mathrm{GeV}/c^2$, applying all the selection 
criteria described above, we find an overall signal selection efficiency of $7.5\%$.

\section{MAXIMUM LIKELIHOOD FIT}
\label{sec:Fit}
The signal content of the data is determined by means of a two-dimensional 
unbinned maximum likelihood fit to the $\dE$ and $\mes$ distributions. 
We consider the following contributions: signal, combinatorial backgrounds 
from continuum and $B\bar{B}$ background processes, $B\to X_d\piz/\eta$ decays, 
$B\to X_s\gamma$ decays, and self-crossfeed from mis-reconstructed  
$B\to X_d\gamma$ decays. 
The likelihood function is defined as
 \begin{equation}
 {\cal L}=\exp{\left(-\sum_{i=1}^{N_{\mathrm{hyp}}} n_{i}\right)}\cdot\left
 [\prod_{j = 1}^{N_{dat}}\left(\sum_{i=1}^{N_{\mathrm{hyp}}} n_i{\cal P}_{i}(\vec{x}_j;\ \vec{\alpha}_i)\right)\right]\; .
 \end{equation}

Here, $n_i$ are the event yields for each of the signal and 
background hypotheses and $N_{dat}$ is the number of events observed in data;
$P_i$ is the probability density function (PDF) for hypothesis $i$, which depends 
on the fit observables $\vec{x}_j=(\dE,\mes)$ and a set of parameters 
$\vec{\alpha}_i$.

The functional form of each PDF is determined from MC simulated events. 
For the \mbox{$\B\to X_d\gamma$} self-crossfeed  component we use 
a binned two-dimensional $(\dE,\mes)$ distribution as the input PDF, while for the 
other hypotheses we obtain the PDFs from unbinned one-dimensional fits to the $\dE$ and $\mes$ 
distributions respectively. For the signal, the $\mes$ spectrum is described by a Crystal 
Ball function~\cite{CryBall}, and the $\dE$ distribution is parametrized as an asymmetric, 
variable-width Gaussian 
\mbox{$f(x) = \exp \left[-(x-\mu)^2/(2 \sigma^2_{L,R} + \alpha_{L,R} 
(x-\mu)^2) \right]$}, where $\mu$ is the peak position of the distribution, 
$\sigma_{L,R}$ are the widths left and right of the peak, and $\alpha_{L,R}$ are measures of the 
tails on the left and right sides of the peak. The contributions from $B\to X_d\piz/\eta$ and 
$\B\to X_s\gamma$ decays are both modeled by adding a Gaussian distribution to an ARGUS 
function~\cite{argus} for $\mes$ 
and a Gaussian distribution to the second-order polynomial describing the background 
for $\dE$. The Gaussian peaks in $\dE$ are
displaced by $\approx 80\mev$ for $B\to X_d\piz/\eta$ by the missing photon, and 
by $\approx 100\mev$ for $\B\to X_s\gamma$ by the kaon mis-identification. 
A combined PDF is defined for the remaining $B$ backgrounds and 
continuum processes using the sum of a Gaussian and an ARGUS function for $\mes$ and 
a second-order polynomial for $\dE$. The Gaussian component allows for 
a small component of $B$ decays that peaks in $\mes$ but not in $\dE$. 

In the likelihood fit the $\mes$ ARGUS slope parameter and the $\dE$ polynomial coefficients of 
the combined background PDF are free parameters, likewise the yields 
for this background and for the signal. All other fit parameters are fixed to the values  
obtained from the MC samples.

\section{VALIDATION OF THE ANALYSIS PROCEDURE}
\label{sec:Validation}

In order to validate the analysis procedure, we embed signal MC events in backgrounds
randomly generated from the background PDFs. The embedded events include both the correctly 
reconstructed and  the self-crossfeed signal contributions. 
For  different numbers of embedded signal events 
 we determine the pulls on the yield and the statistical error from the fit. 
We find a bias in the fitted yield of +24\% of the statistical error.
We correct the final fit result for this bias and treat it as a systematic
error.

To check our analysis on data we select $\B\to K^{*}\g$ decays, using the $K^*$ decay modes 
$\Kstarp \to \Kp \piz$ and $\Kstarz \to \Kp \pim$. The selection is the same as for the two signal modes 
\itypeone\ and \itypetwo, except that the \pipm\ identification requirements are replaced with 
a charged kaon selection, and the mass range is $0.6\gevcc<M(X_s)<1.0\gevcc$. In the fit procedure described in the 
previous section we include a $B\to X_s\gamma$ self-crossfeed component and a $B\to X_s\piz/\eta$ background, 
but not a $B\to X_d\gamma$ misidentification component because this is expected to be negligible. 
$\B\to K^{*}\g$ has a large signal yield, so we use this fit to determine the signal shape in data 
by allowing the means and widths of the signal PDFs to vary.
We find $1680\pm51$ $\B\to K^{*}\g$ events in the \babar\ data, where this error is purely statistical. 
This is in excellent agreement with the expectation of $1616\pm28$ events based on the previously measured  
branching fraction of $\BR(\B\to K^{*}\g) = (4.02 \pm0.33)\times 10^{-5}$~\cite{hfag}.

\begin{figure}[htb]
\begin{center}
\hspace*{-0.4cm}
\includegraphics[width=0.54\textwidth]{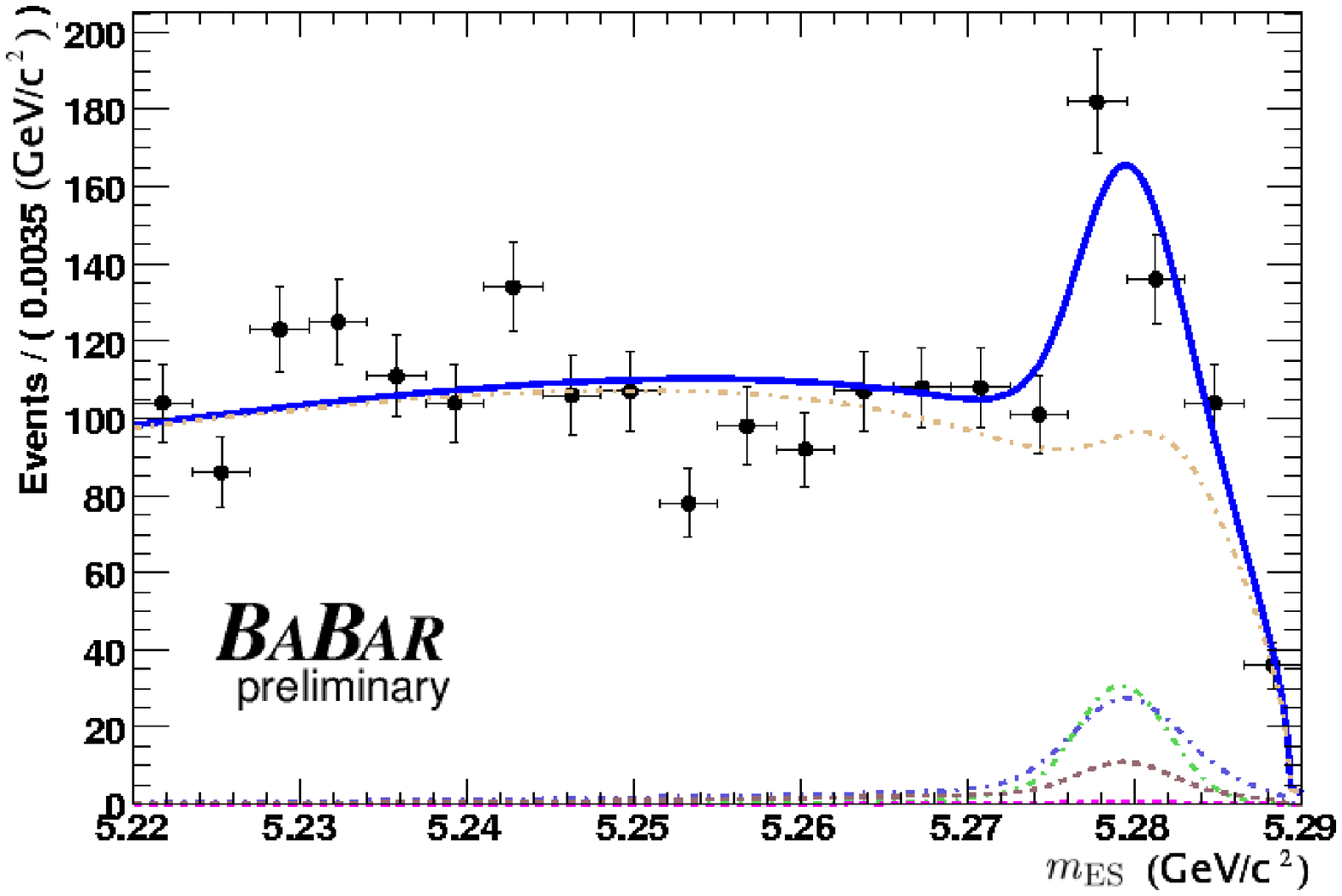}\hspace*{-0.2cm} 
\includegraphics[width=0.54\textwidth]{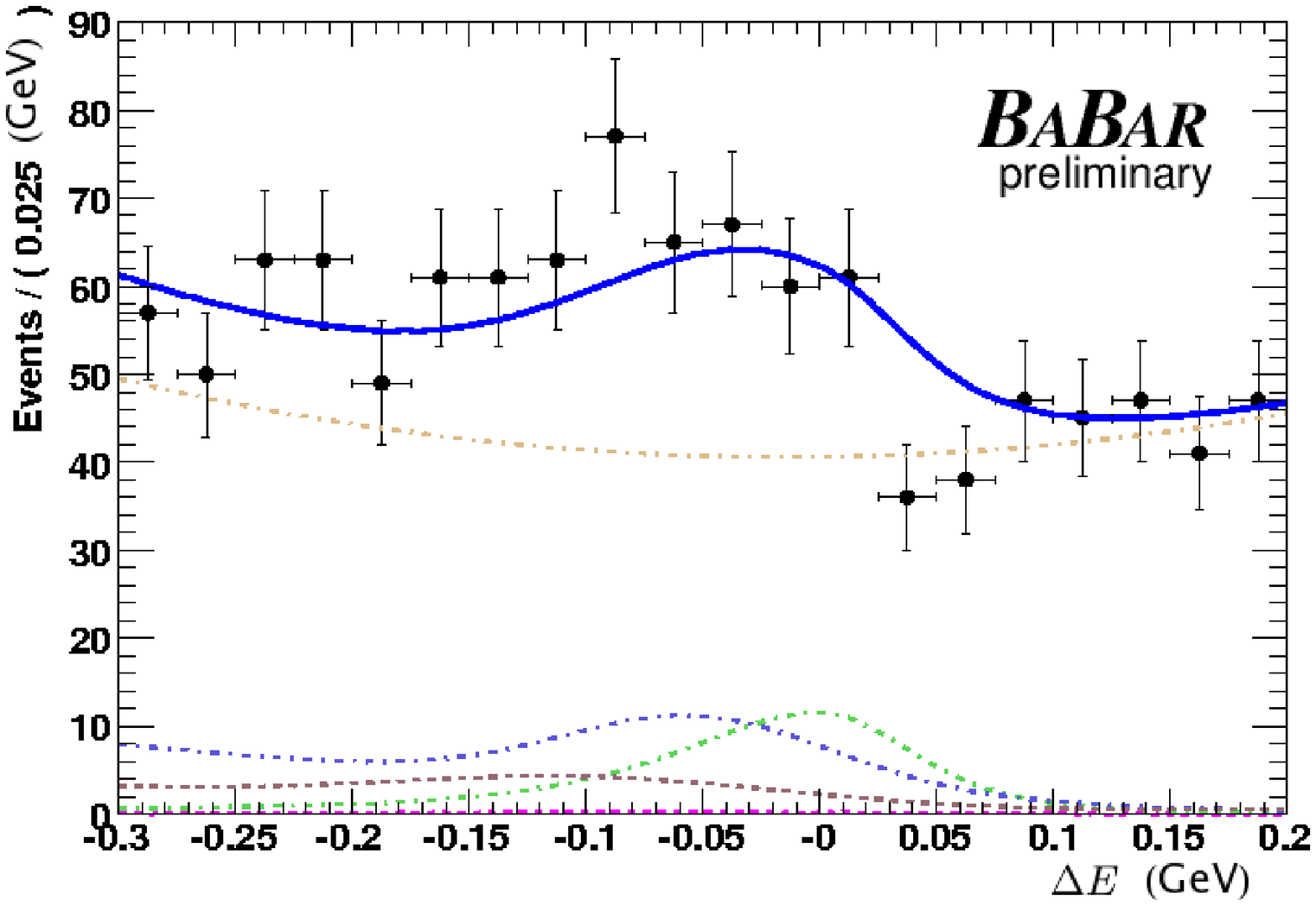}
\caption{Projections of 
$\mes$ with $-0.1\gev<\de<0.05\gev$ (left) and $\dE$ with $5.275\gevcc<\mes<5.286\gevcc$ (right) 
in the $B\to(\rho,\omega)\gamma$ fit to data (points with statistical errors).   
The curves represent the PDFs used in the fit with the combined PDF in solid blue. The dashed lines represent the individual PDFs  with with signal component in green, signal cross-feed in pink, \btosgam\ background in blue, $B\to X_d\piz/\eta$  background in brown and background in orange.  }

\label{fig:rho-proj} 
\end{center}
\vspace*{-2ex}
\end{figure}

As a further check on data we perform the signal analysis for the decays \itypeone, \itypetwo, and \itypefour\ in the  
hadronic mass range $0.6\gevcc<M(X_d)<1.0\gevcc$, which contains the $\rho^0, \rho^{\pm}$ and $\omega$ resonances. 
In this fit we include all the contributions in a similar way to the $B\to X_d\gamma$ analysis 
in the higher mass region, and fix the signal PDF shape to the values obtained from the $\B\to K^{*}\g$ fit.
We obtain a combined $B\to(\rho,\omega)\gamma$ signal of $73\pm25$(stat.) events corresponding to a 
statistical significance of $2.9\sigma$. This is consistent with the $66\pm26$ events expected from 
the average of previous branching fraction measurements by \babar\ and Belle of 
$\BR(\B\to (\rho,\omega)\gamma) = (1.28 \pm0.21)\times 10^{-6}$~\cite{hfag}. Note that we do not  
make use of the resonance mass or the decay helicity angle to discriminate against backgrounds, so  
we do not expect this measurement to be as precise as the exclusive $B\to(\rho,\omega)\gamma$ analyses.
Figure~\ref{fig:rho-proj} shows a comparison of the PDF component shapes (curves) to the data (points). 

\section{RESULTS}
\label{sec:Results}

Finally we perform the full fit for the sum of all seven decay modes in the 
hadronic mass range $1.0\gevcc<M(X_d)<1.8\,\gevcc$, again using 
the signal PDFs from $\B\to K^{*}\g$.

\begin{figure}[bht]
\begin{center}
\hspace*{-0.4cm}
\includegraphics[width=0.54\textwidth]{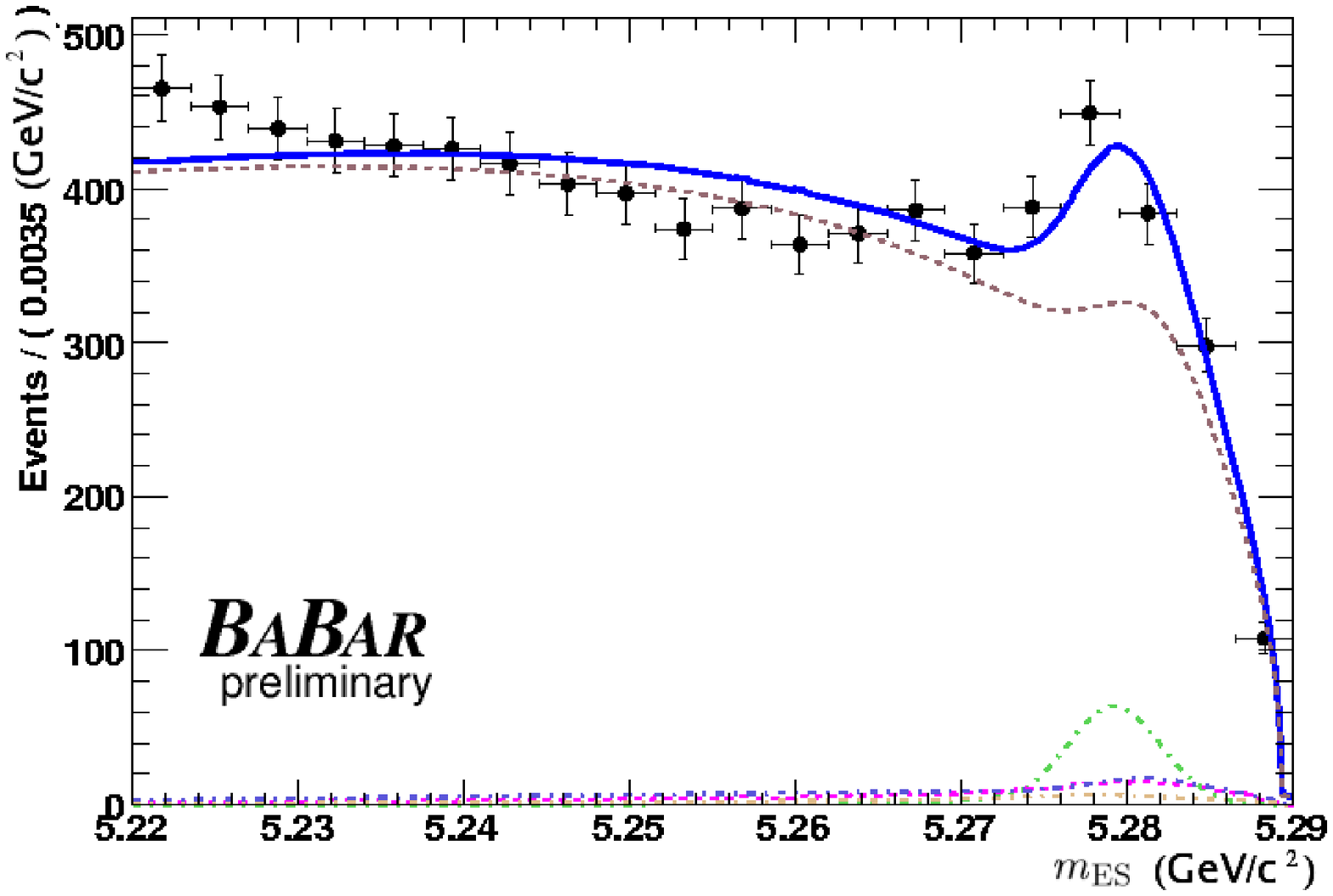}\hspace*{-0.2cm} 
\includegraphics[width=0.54\textwidth]{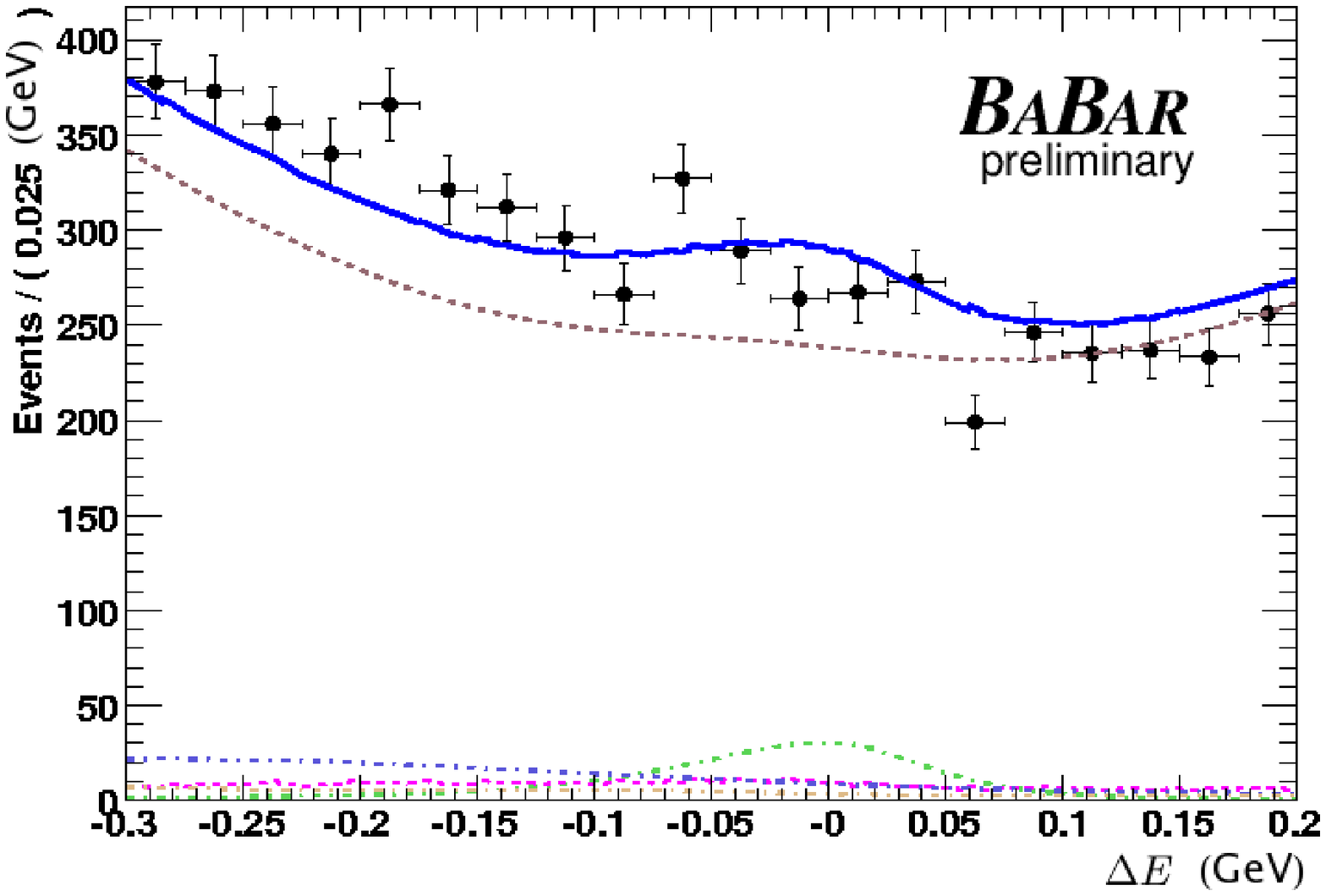}
\caption{Projections of 
$\mes$ with $-0.1\gev<\de<0.05\gev$ (left) and $\dE$ with $5.275\gevcc<\mes<5.286\gevcc$ (right) 
 in the \btodgam\ fit to data.  
The curves represent the PDFs used in the fit with the combined PDF in solid blue. The dashed lines represent the individual PDFs  with with signal component in green, signal cross-feed in pink, \btosgam\ background in blue, $B\to X_d\piz/\eta$  background in orange and background in brown.}

\label{fig:bdg-proj} 
\end{center}
\vspace*{-2ex}
\end{figure}

Figure~\ref{fig:bdg-proj} shows a comparison of the PDF component shapes (curves) to the data (points). 
For each plot, a cut is applied to the variable not plotted around the signal peak. 

\begin{figure}[!t]
\begin{center}
\hspace*{-0.4cm}
\includegraphics[width=0.54\textwidth]{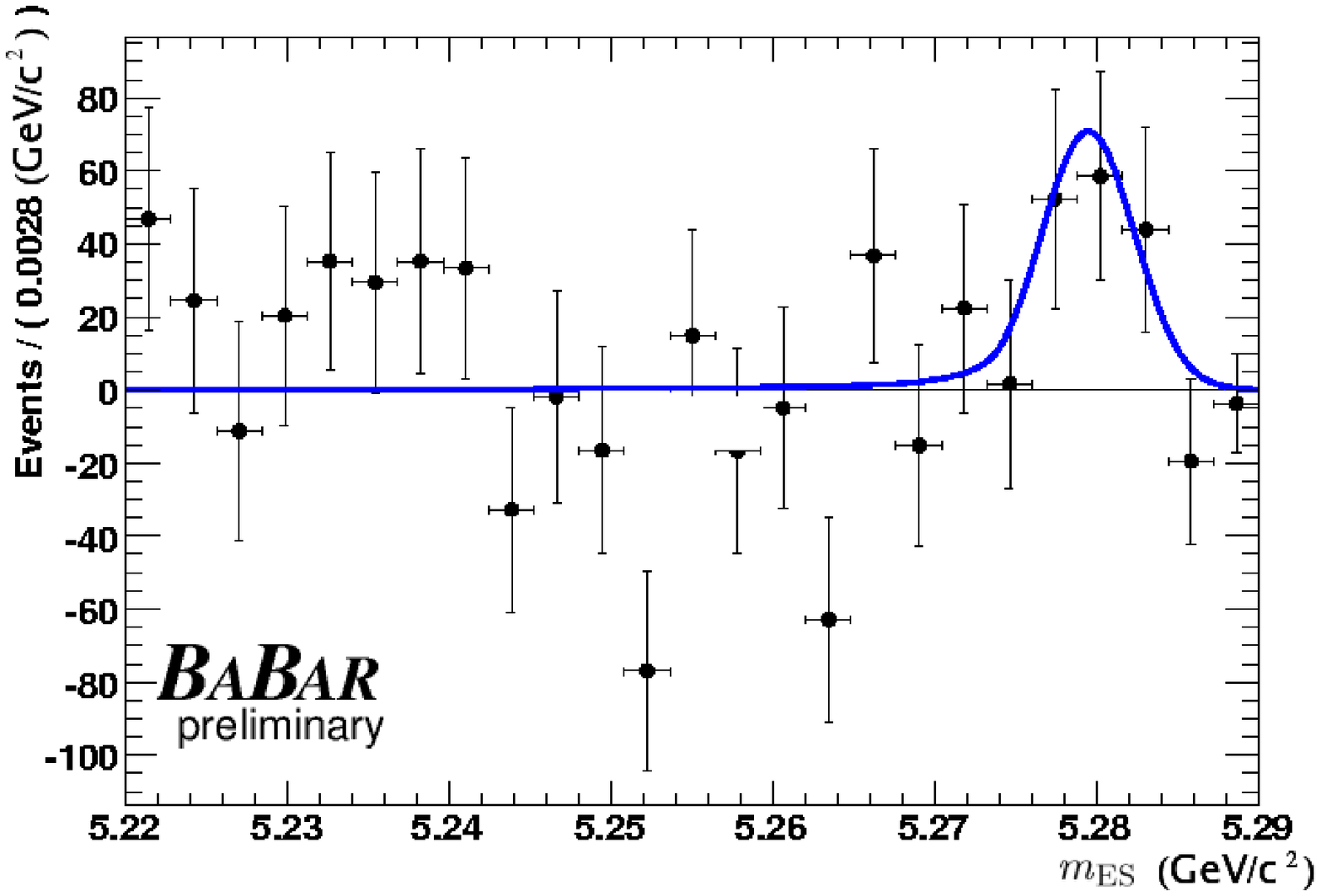}\hspace*{-0.2cm}
\includegraphics[width=0.54\textwidth]{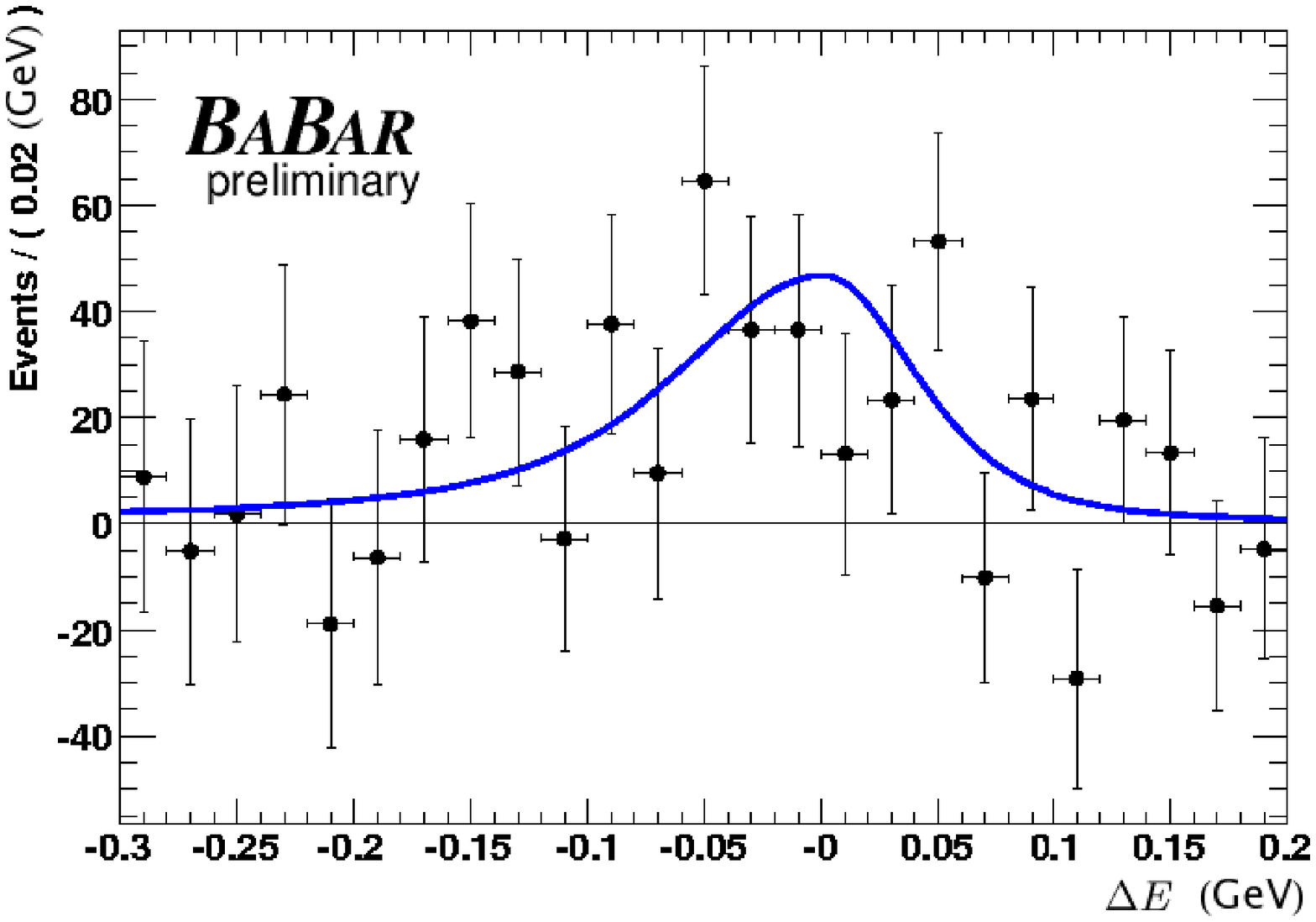}
\caption{Signal 
distributions of 
$\mes$ (left) and $\dE$ (right) for $B \to \X_d \gamma$ in the range $1.0\gevcc<M(X_d)<1.8\gevcc$
with the background subtraction as described in the text.  The curves
represent the PDFs used in the fit, normalized to the fitted yield.}
\label{fig:fit} 
\end{center}
\vspace*{-2ex}
\end{figure}

Figure~\ref{fig:fit} shows a comparison of the PDF shapes (solid curves) to the data using 
the event-weighting technique  described in Ref.~\cite{sPlots} to subtract the fitted background. 
For each plot, we perform 
a fit excluding the variable being plotted and use the fitted yields and covariance matrix 
to determine the relative probability that an event is signal or background.  The distribution 
is normalized to the yield for the given component and can be compared directly to the assumed 
PDF shape. We find good agreement between the data and the PDFs.

We find a signal yield of $178\pm 53$ events and a background of $27670\pm173$ events in the full fit. 
Taking only 
statistical uncertainties into account, this corresponds to a signal significance of $3.4\sigma$. 
We calculate the partial branching fraction summed over the seven modes in the mass region  
$1.0\gevcc<M(X_d)<1.8\,\mathrm{GeV}/c^2$ using the following formula: 
\begin{equation}
\sum\limits^{7}_{X_d=1}\BR(B\to X_d\gamma)|_{(1.0\gevcc<M(X_d)<1.8\,\mathrm{GeV}/c^2)}=
\frac{n_{sig}} {2\ \epsilon\  n_{B\overline{B}}}
 ; 
\end{equation} 
where $n_{sig}$ is the number of fitted signal events, $\epsilon$ is the signal selection efficiency, $n_{B\overline{B}}$ is the number of $B\overline{B}$ pairs in the dataset, and the factor 2 is to account for the possibility of either $B$ decaying into a signal mode.  
We obtain the following result: 
\begin{equation}
\sum\limits^{7}_{X_d=1}\BR(B\to X_d\gamma)|_{(1.0\gevcc<M(X_d)<1.8\,\mathrm{GeV}/c^2)}=
[3.1\pm 0.9 (stat.)]\cdot 10^{-6}\, ; 
\end{equation}
The relevant systematic uncertainties are discussed below.

\section{SYSTEMATIC UNCERTAINTIES}
\label{sec:Systematics}
Table~\ref{tab:syst} gives an overview of the contributions
to the systematic uncertainties. These are associated with the signal
reconstruction efficiency, the modeling of the signal and $\BB$ background
PDFs in the likelihood fit, and the modeling of the $B\to X_d\gamma$ spectrum and 
the fragmentation of the $X_d$ system. 
\begin{table}[htb]
\vspace*{0ex}
\renewcommand{\arraystretch}{1.3}
\centering
\caption{\label{tab:syst}
Fractional systematic errors of the measured branching fraction.}
\vspace*{1ex}
\begin{tabular}{|l|c|}
\hline
Source & \multicolumn{1}{c|}{Relative uncertainty (\%)}  \\
\hline\hline
Tracking efficiency                 & 1.7    \\ 
Charged-particle identification     & 2.0     \\ 
Photon selection                    & 2.5     \\ 
$\pi^0$ and $\eta$ reconstruction   & 1.7     \\ 
$\pi^0$ and $\eta$ veto             & 1.0     \\ 
NN efficiency                       & 5.0     \\ 
PDF shapes                   	    & 13.0       \\ 
$B$ background normalization        & ${}^{+10.5}_{-3.0}$    \\ 
Fit bias                            & 7.1     \\ 
$B$ counting                        & 1.1     \\ 
\hline
Combined                            & $^{+19.3}_{-16.5}$    \\
\hline \hline

Signal model                        & 3.4      \\
$X_d$ fragmentation model           & 13.6     \\
\hline
Combined                            & 14.7    \\
\hline
\end{tabular}
\vspace*{-2ex}
\end{table}

The systematic errors affecting the signal efficiency includes uncertainties on tracking, 
particle identification, \g\ and \piz\ reconstruction, the \piz/$\eta$ veto, and the neural 
network selection. These uncertainties are evaluated using independent data and 
MC-simulated event samples. They give a combined systematic error of 6.5\% when added in quadrature.  

To estimate the uncertainties related to the signal and background 
yields we vary the parameters of the PDFs that are fixed in the fit.  
For the signal PDF the means and widths are varied within the range allowed 
by the fit to the $\B\to K^{*}\g$ data. 
Other fixed PDF parameters are fluctuated within their statistical errors. 
The combination gives errors of $\pm 13\%$ on the signal yield.

All relative and absolute normalizations of background 
components that are fixed in the fit are also varied. The absolute normalizations of 
the $\B\to X_d\gamma$ cross feed, $\B\to X_s\gamma$, and $B\to X_d\piz/\eta$ components 
are changed by $\pm50\%$,  $\pm20\%$, and $\pm100\%$ respectively; and the relative contribution 
of the remaining $B$ decays to the combined background component is varied by  $\pm20\%$.
The full size of the bias on the signal yield observed  
in MC experiments (see Section~\ref{sec:Validation}), is taken as a systematic uncertainty 
of the fit procedure. 

There is a small 1.1\% uncertainty on the overall normalization associated with the 
the total number of $\BB$ pairs in the underlying data sample.

The impact of the assumed $X_d$ spectrum on the efficiency is studied by using two different 
sets of values for the kinetic parameters of the Kagan-Neubert model~\cite{KaganNeubert} 
$(m_b,\mu_{\pi}^2) = (4.65, -0.50)$ and $(4.80, -0.35)$. These are consistent 
with fits of $B\to X_s\gamma$ data~\cite{OliverHenning}.  
The default fragmentation model for the $X_d$ is JETSET~\cite{JETSET}, 
 known not to give a good description of the distribution of final states in 
$B\to X_s\gamma$~\cite{xsgam}. To estimate the uncertainty associated with the fragmentation model, 
we re-weight the relative contributions of the seven signal modes according  to results obtained in 
\cite{xsgam} for the corresponding $B\to X_s\gamma$ decays, and take the resulting change in the signal 
yield as a systematic error. 
 Combining these two uncertainties, we assign a systematic error of 14.7/
the modeling of the signal.

\section{SUMMARY}
\label{sec:Summary}

In this paper we have demonstrated the feasibility of measuring  $B\to X_d\gamma$ decays 
with a semi-inclusive technique using a sum of seven exclusive final state in the mass range 
$1.0\gevcc<M(X_d)<1.8\,\mathrm{GeV}/c^2$ and presented the first evidence for  
$b\to d\gamma$ transitions in this mass range. 
We find $178\pm 53$ signal events. 

We measure a partial branching fraction 
$\sum\nolimits^{7}_{X_d=1}\BR(B\to X_d\gamma)|_{(1.0\gevcc<M(X_d)<1.8\,\mathrm{GeV}/c^2)}= 
(3.1\pm0.9 ^{+0.6}_{-0.5} \pm0.5)\cdot 10^{-6},$ 
where the uncertainties are statistical, systematic and model respectively. 
Including statistical and systematic 
uncertainties only, this corresponds to a signal significance of $3.1\sigma$.

\section{OUTLOOK}
\label{sec:Outlook}
In the future, the inclusive \btodgam\ transition rate can be measured. 
This will require further study to:  
\begin{itemize}
\item Extend the current measurement to the mass region $0.6\gevcc<M(X_d)< 1.0\gevcc$,  
which has been used as a control sample in this analysis with only three decay 
modes reconstructed. 
We can measure the seven decay modes over both ranges. 

\item Correct for the part of the $X_d$ fragmentation that is not reconstructed 
by the seven decay modes. 
Based on MC simulation and the JETSET model we expect the measured seven modes to account for about 50\%\ of the the hadronic mass region $1.0\gevcc<M(X_d)<1.8\gevcc$. 

\item Extrapolate to the full mass region. Assuming the shape of the hadronic mass spectrum is the same for \btodgam\  as for \btosgam\ decays \cite{xsgam}, the mass region $0.6\gevcc - 1.8\gevcc$ comprises roughly 60\%\ of the spectrum. 
\end{itemize}

Further, the extraction of the ratio $|V_{td}/V_{ts}|$ can be performed by comparing the branching 
fractions for $\btodgam$ and $\btosgam$ over the experimentally accessible 
final states and hadronic mass range.  
By measuring this ratio over a larger hadronic mass range, and with a larger set of 
final states it should be possible to reduce the theoretical uncertainties compared 
to the ratio of $B\to \rho\gamma$ and $B\to K^*\gamma$ decays.

\section{ACKNOWLEDGMENTS}
\label{sec:Acknowledgments}

We are grateful for the 
extraordinary contributions of our \pep2\ colleagues in
achieving the excellent luminosity and machine conditions
that have made this work possible.
The success of this project also relies critically on the 
expertise and dedication of the computing organizations that 
support \babar.
The collaborating institutions wish to thank 
SLAC for its support and the kind hospitality extended to them. 
This work is supported by the
US Department of Energy
and National Science Foundation, the
Natural Sciences and Engineering Research Council (Canada),
the Commissariat \`a l'Energie Atomique and
Institut National de Physique Nucl\'eaire et de Physique des Particules
(France), the
Bundesministerium f\"ur Bildung und Forschung and
Deutsche Forschungsgemeinschaft
(Germany), the
Istituto Nazionale di Fisica Nucleare (Italy),
the Foundation for Fundamental Research on Matter (The Netherlands),
the Research Council of Norway, the
Ministry of Science and Technology of the Russian Federation, 
Ministerio de Educaci\'on y Ciencia (Spain), and the
Science and Technology Facilities Council (United Kingdom).
Individuals have received support from 
the Marie-Curie IEF program (European Union) and
the A. P. Sloan Foundation.


\begin{thebibliography}{99}
\bibitem{conj} Charge conjugate states are implied throughout the paper.
\end{thebibliography}
\end{document}